\documentclass[superscriptaddress,twocolumn,showpacs,preprintnumbers,amsmath,amssymb]{revtex4}

\usepackage{graphicx}
\usepackage{dcolumn}
\usepackage{bm}
\usepackage{amsmath}
\usepackage{epsfig}
\usepackage{rotating}
\usepackage{enumerate}
\usepackage{xcolor}
\usepackage{longtable}

\begin{document}

\title{Descriptors of intrinsic hydrodynamic thermal transport:\\ screening a phonon database in a machine learning approach}

\author{Pol Torres}
\affiliation{Department of Mechanical Engineering, The University of Tokyo, 7-3-1 Hongo, Bunkyo,
Tokyo, 113-8656, Japan}
\affiliation{EURECAT, Technology Center of Catalonia, Applied Artificial Intelligence, 08290 Cerdanyola, Barcelona, Spain}
\affiliation{Departament de F\'isica, Universitat Aut\`onoma de Barcelona (UAB), Campus de Bellaterra, 08193 Bellaterra, Barcelona, Spain}

\author{Stephen Wu}
\affiliation{Research Organization of Information and Systems, The Institute of Statistical Mathematics
(ISM), 10-3 Midori-cho, Tachikawa, Tokyo 190-8562, Japan}

\author{Shenghong Ju}
\affiliation{China-UK Low Carbon Collage, Shanghai Jiao Tong University, Shanghai 201306, China}
\affiliation{Department of Mechanical Engineering, The University of Tokyo, 7-3-1 Hongo, Bunkyo,
Tokyo, 113-8656, Japan}

\author{Chang Liu }
\affiliation{Center for Materials Research by Information Integration (CMI2), Research and Services
Division of Materials Data and Integrated System (MaDIS), National Institute for Materials
Science (NIMS), 1-2-1 Sengen, Tsukuba, Ibaraki 305-0047, Japan}
\affiliation{Research Organization of Information and Systems, The Institute of Statistical Mathematics
(ISM), 10-3 Midori-cho, Tachikawa, Tokyo 190-8562, Japan}

\author{Terumasa Tadano}
\affiliation{Research Center for Magnetic and Spintronic Materials, National Institute for Materials and Science, Tsukuba, Japan}

\author{Ryo Yoshida}
\affiliation{Research Organization of Information and Systems, The Institute of Statistical Mathematics
(ISM), 10-3 Midori-cho, Tachikawa, Tokyo 190-8562, Japan}
\affiliation{Center for Materials Research by Information Integration (CMI2), Research and Services
Division of Materials Data and Integrated System (MaDIS), National Institute for Materials
Science (NIMS), 1-2-1 Sengen, Tsukuba, Ibaraki 305-0047, Japan}

\author{Junichiro Shiomi}
\affiliation{Department of Mechanical Engineering, The University of Tokyo, 7-3-1 Hongo, Bunkyo,
Tokyo, 113-8656, Japan}
\email{shiomi@photon.t.u-tokyo.ac.jp}

\date{\today}

\begin{abstract}

Machine learning techniques are used to explore the intrinsic origins of the hydrodynamic thermal transport and to find new materials interesting for science and engineering. The hydrodynamic thermal transport is governed intrinsically by the hydrodynamic scale and the thermal conductivity. The correlations between these intrinsic properties and harmonic and anharmonic properties, and a large number of compositional (290) and structural (1224) descriptors of 131 crystal compound materials are obtained, revealing some of the key descriptors that determines the magnitude of the intrinsic hydrodynamic effects, most of them related with the phonon relaxation times. Then, a trained black-box model is applied to screen more than 5000 materials. The results identify materials with potential technological applications. Understanding the properties correlated to hydrodynamic thermal transport can help to find new thermoelectric materials and on the design of new materials to ease the heat dissipation in electronic devices.

\end{abstract}

\maketitle

\section{INTRODUCTION}

From the Boltzmann transport equation (BTE) for phonons it is well known that the thermal conductivity of a crystalline material is mainly determined by the phonon scattering. The stronger the scattering is, the higher the thermal resistance is. Phonons can scatter with each other anharmonically, against boundaries, with impurities and defects. The BTE states that when the phonon distribution is perturbed from its equilibrium, it relaxes to the equilibrium through the scattering events~\cite{Guyer1966a, Esfarjani2008, Chaput2013, Torres_book}.

When studying the thermal transport in bulk materials and under slow heating conditions, the Fourier's law provides a simple expression for the diffusive thermal transport:

\begin{equation}\label{eq_Fourier}
\mathbf{Q}=-\mathbf{\kappa}\pmb{\nabla}T \; ,
\end{equation}
where $\mathbf{Q}$ is the heat flux, $\mathbf{\kappa}$ the thermal conductivity and $T$ the temperature. In recent experiments at the micro/nano scale and/or using fast and/or large thermal gradient, it has been observed that not all thermal transport can be explained by the Fourier's law~\cite{Johnson2013, Siemens2010a, Minnich2011, Regner2013, Wilson2014, Hu2015, Hoogeboom-Pot2015}. In such situations, other transport phenomena such as ballistic transport and L\'evy flights~\cite{Vermeersch2014,Vermeersch2014b} or hydrodynamic transport appear~\cite{Cepellotti2015, Lee2015, Torres2018a}. In this article we will focus on the latter transport regime. 

In the second half of the last century, Guyer and Krumhansl introduced an expression of the thermal transport beyond the Fourier's law to describe the hydrodynamic thermal transport\cite{Guyer1966a, Guyer1966}:

\begin{equation}\label{c1_eq_gk}
\tau \dot{\textbf{Q}}+\textbf{Q}=-\kappa \pmb{\nabla}T+\ell^2 \nabla^2 \textbf{Q} \; ,
\end{equation}
where $\tau$ is the phonon mean free time and $\ell$ is the hydrodynamic scale. Both of these are the values averaged over the entire phonon spectrum~\cite{Torres_book, Torres2018a}. Eq.~(\ref{c1_eq_gk}) is a general hydrodynamic heat flux equation that includes memory (time derivative term) and non-local effects (Laplacian term), and it has been used successfully to describe recent experiments~\citep{Amir2017, Torres2018a} that could not be reproduced only by Fourier's law.

It is known that the thermal hydrodynamic effects are related with the relative importance of Normal (N) scattering (where phonon-phonon collisions conserve momentum) in front of restive (R) scattering (where phonon-phonon collisions do not conserve momentum, like Uumklapp scattering or mass deffect), but little is known about the intrinsic origin of these phenomena. In addition, as described by Eq.~(\ref{c1_eq_gk}), the temporal scale and the geometry will also contribute to this effect. To elucidate the key structural/chemical factors that determine the hydrodynamic thermal transport, leaving aside the temporal and geometrical scales, in this paper we employ machine learning techniques. A Python open source platform of materials informatics that we have developed called XenonPy~\cite{xenonpy} is used to generate compositional and sctructural descriptors. By providing only the Materials Project ID of a certain material~\cite{Jain2013, ONG2015209}, the tool provides 290 compositional descriptors and 1224 structural descriptors like the polarizability, lattice constant, and Van Der Waals radius, etc.

\section{METHODOLOGY}

In this work we perform full first-principles calculations for 131 materials. The interatomic force constants from 94 of those materials are provided by A. Togo~\cite{PhysRevLett.115.205901}, 18 are from the AlmaBTE database~\cite{PhysRevX.4.011019} and 20 are calculated by ourselves. Our DFT calculations are performed using the Vienna Ab initio Simulation Package (VASP) in the framework of the projector augmented
wave (PAW) potentials in the Perdew-Burke-Ernzerhof (PBE) approximation. The energy cutoff was fixed at a value 10\% higher than the value specified in the pseudopotential file. The atom displacements to calculate the harmonic and anharmonic interatomic force constants are provided by the Phonopy~\cite{phonopy} and Phono3py~\cite{phono3py} codes respectively.

The parameters that describe the hydrodynamic thermal transport from Eq.~(\ref{c1_eq_gk}), mainly $\kappa$, $\tau$ and $\ell$ are computed from the first-principles calculations beyond the relaxation-time approximation (RTA) by using Phono3py and the KCM code~\cite{Torres2017}. All the calculations are performed at 300~K. From the calculations, we have obtained extra first-principles descriptors like the averaged Normal and resistive (Umklapp plus mass variance/impurity) relaxation times ($\tau_N$, $\tau_R$), specific heat $C$, and the group velocity $v$. In recent articles, it has been described that other parameters such as the maximum frequency of the phonon dispersion relations ($\omega_{max}$), the Gr\"uneisen parameter, the volume of the unit cell ($V$), and the scattering phase space ($SPS$) show a good correlation with the thermal conductivity~\cite{Abhishek2019, Shenghong2020}. These are harmonic and quasi-harmonic properties but not all of them are equally easy to calculate. While the maximum frequency is obtained with a single harmonic calculation, evaluation of the Gr\"uneisen parameter requires a few calculations with different volumes. The $SPS$ is defined as:

\begin{equation}\label{eq_SPS}
P_3^ \pm  (\mathbf{q}j) = \frac{1}{N_{\mathbf{q}}} \sum_{q1,q2,j1,j2} \delta (\omega_{q,j} \pm \omega_{q1,j1}-\omega_{q2,j2})\delta_{q\pm q1,q2+G} \; ,
\end{equation}
where $\omega$ and $\mathbf{q}$ are the phonon frequency and the crystalline momentum of each phonon involved in the three-phonon interaction. Therefore, although $SPS$ is a harmonic property, it requires exploration of all the three-phonon scattering processes, which takes a some time.

The recent works have found good correlation between $w_{max}$ and $\kappa$~\cite{Abhishek2019, Shenghong2020}, and from our experience we know that in several cases $\kappa$ shows a good correlation with $\ell$. The correlation of $w_{max}$ and $\kappa$ can be broken in materials with high frequency optical bands. Those bands are usually flat and therefore the velocity is small and does no contribute significantly to the thermal conductivity. For instance, lithium hydride (LiH) and lithium fluoride (LiF) have $\omega_{max} \sim 32$~THz ($\sim 20$~THz) and $\kappa = 22$~W/mK ($14$~W/mK) respectively~\cite{PhysRevB.94.174304} while silicon (Si) has $\omega_{max} \sim 16$~THz and $\kappa = 145$~W/mK at 300~K~\cite{Ward2010}.

In the effort of finding a new descriptor including the velocity to avoid the effect of high frequency optical bands, we propose two candidates. The first one is equivalent to the Landauer thermal conductance expressed in the $\mathbf{q}$ space considering the transmission $T=1$ for all the modes:
\begin{equation}
G_{max} = \frac{1}{N_{\mathbf{q}} V}\sum_{\mathbf{q},j} \frac{\hbar \omega}{2 \pi k_B T^2}  \frac{e^{\hbar \omega/k_B T}}{(e^{\hbar \omega/k_B T}-1)^2} \textbf{v} \; ,
\label{land_q}
\end{equation}
where $N$ and $V$ are number of {\bf q}-points in the mesh sampling and the volume of the unit cell, respectively. The second descriptor is a pseudo thermal conductivity defined in the following. We know that the thermal conductivity under the RTA can be expressed as:
\begin{equation}
\kappa^{\alpha\beta} = \frac{1}{N_{\mathbf{q}} V} \sum_{\mathbf{q},j} C_{\mathbf{q},j} v_{\mathbf{q},j}^{\alpha}v_{\mathbf{q},j}^{\beta}\tau_{\mathbf{q},j} \; ,
\label{kappa_rta}
\end{equation}
where $\alpha, \beta$ are cartesian cordinates, $C_{\mathbf{q},j}$ the specific heat and $\tau_{\mathbf{q},j}$ the relaxation time. The latter is calculated from the phonon-phonon scattering rate as $\tau_{\mathbf{q},j}=\hbar/2\Gamma$, where
\begin{equation}
\begin{aligned}
\Gamma_{\mathbf{q}j} = \frac{\pi \hbar}{2 N_{\mathbf{q}j}} \sum_{\mathbf{q}'j'\mathbf{q}''j''} \frac{|\Phi (\mathbf{q}j\mathbf{q}'j'\mathbf{q}''j'')|^2}{8\omega_{\mathbf{q}j}\omega_{\mathbf{q'}j'}\omega_{\mathbf{q''}j''}} \cdot \\
[(1+n_{\mathbf{q}'j'}+n_{\mathbf{q}''j''})\delta(\omega_{\mathbf{q}j}-\omega_{\mathbf{q}'j'}-\omega_{\mathbf{q}''j''}) \\
+ 2(n_{\mathbf{q}'j'}-n_{\mathbf{q}''j''})\delta(\omega_{\mathbf{q}j}+\omega_{\mathbf{q}'j'}-\omega_{\mathbf{q}''j''})] \; ,
\end{aligned}
\label{eq_gamma}
\end{equation}
and $|\Phi (\mathbf{q}\mathbf{q}'\mathbf{q}'')|^2$ is the three-phonon scattering matrix. If we approximate $|\Phi (\mathbf{q}\mathbf{q}'\mathbf{q}'')|^2/8\omega_{\mathbf{q}j}\omega_{\mathbf{q'}j'}\omega_{\mathbf{q''}j''} \rightarrow 1/m^2$, then
\begin{equation}
\begin{aligned}
\Gamma_{\mathbf{q}} = \frac{\pi\hbar}{2 m^2} (W^-_{\mathbf{q},j}+2W^+_{\mathbf{q},j}) \; .
\end{aligned}
\label{eq_gamma2}
\end{equation}
The magnitude $W^{\pm}_{\mathbf{q},j}$ is the temperature dependent $SPS$ ($TSPS$), which as well as the $SPS$, can be obtanied directly from Phono3py~\cite{phono3py}. With these definitions, we introduce the pseudo thermal conductivity as:
\begin{equation}
\kappa^{\alpha\beta}_{pseudo} = \frac{1}{N_{\mathbf{q}} V} \sum_{\mathbf{q},j} C_{\mathbf{q},j} v_{\mathbf{q},j}^{\alpha}v_{\mathbf{q},j}^{\beta} \left[ \frac{\pi}{m^2}(W^-_{\mathbf{q},j}+2W^+_{\mathbf{q},j}) \right]^{-1} \; .
\label{kappa_pseudo}
\end{equation}

\section{RESULTS AND DISCUSSION}

The values of $\ell$, $\kappa$ and all the descriptors described in the previous section has been calculated for the 131 materials from first-principles and with XenonPy. 

The first-principles calculations show that the materials with stronger intrinsic hydrodynamic behaviour (larger $\ell$) are those made of carbon, such as diamond (mp-66), lonsdaleite (mp-47), and other carbon allotropes (mp-611426 and mp-616440), as well as  BAs in wurtzite(WZ) and zinc-blend(ZB) structures. The former carbon based materials have a thermal conductivity of around 2500~W/mK while the hydrodynamic scale ranges from 800-1000~nm. In comparison with BAs, the $\kappa$ is smaller while $\ell$ is larger. More explicity, WZ(ZB) BAs have a thermal conductivity of $\sim$1250~W/mK($\sim$1500~W/mK) and $\ell \sim$1100~nm($\sim$1400~nm). Another case is BN, which has a large thermal conductivity (1200~W/mK) but smaller $\ell$ (319~nm) than BAs in wurtzite (WZ) and zinc-blend (ZB) phases. Those materials with large $\ell$ are interesting from the viewpoint of studying the experimental signatures of the hydrodynamic thermal transport, as they also exhibit a high relative importance of N versus resistive scattering processes, as reported in several works~\citep{Ward2009, Broido2013a}. Also, these materials can be usefull to study the improvement of heat release in hot spots in electronic devices due to fast operation rates, as hydrodynamic effects near to the edges of electronic circuits create a vorticity that helps to increase the heat dissipation due to an effective change of the thermal conductivity~\cite{Torres2018a}.

\begin{table}[h!] 
\begin{center}
\caption{Materials with $\ell > \gamma \kappa$. $\kappa$ is expressed in W/mK and $\ell$ in nm. More materials fulfilling this condition can be found in the APPENDIX.}\label{table_low}
\begin{tabular}{lcccccc}
\hline
\textbf{Material} & $\mathbf{\kappa}$ & $\mathbf{\ell}$ & & \textbf{Material} & $\mathbf{\kappa}$ & $\mathbf{\ell}$\\
\hline
AlP(mp-1550) & 101.5 & 151.7 && GaN(mp-830) & 201.4 & 202.3\\
\hline
BeS(mp-422) & 164.4 & 173.2 && BeTe(mp-252) & 200 & 351.7\\
\hline
Bi$_2$Te$_3$ & 1.5 & 10 && NaBr(mp-22916) & 3.4 & 11.6\\
(mp-1227339) & & && & & \\
\hline
KBr(mp-23251) & 3.3 & 17.1 && BaSe(mp-1253) & 11.5 & 53.2\\
\hline
BaTe(mp-1000) & 13.4 & 63.8 && AgI(mp-22925) & 1.2 & 42.6\\
\hline
CdS(mp-2469) & 25.6 & 111.3 && CdSe(mp-2691) & 8.2 & 59.4\\
\hline
CdTe(mp-406) & 6 & 42.6 && CuBr(mp-22913) & 1 & 11.2\\
\hline
GaAs(mp-2534)  & 36.5 & 94.3 && GaP(mp-2490) &  84.7 & 180.0 \\
\hline
GaSb(mp-1156) &  40.8 & 145.4 && InAs  & 24.6 & 97.0 \\
& & && (mp-20305) & & \\
\hline
InP(mp-20351) &  77.0 &  426.2 && InSb & 15.6 & 77.8 \\
& & && (mp-20012) & & \\
\hline
MgTe(mp-13033) & 15.8 & 88.8 && ZnS &  23.4 & 75.6 \\
& & && (mp-10695) & & \\
\hline
ZnSe(mp-1190) & 17.5 & 52.4 && ZnTe(mp-2176) & 15.9 & 96.2 \\
\hline
AgI(mp-22894) & 1 & 18.6 && AlAs(mp-8881) &  82.3 & 198.6 \\
\hline
AlP(mp-8880)  & 83.1 & 126.5 && AlSb & 84.7 & 302  \\
& & && (mp-1018100) & & \\
\hline
CdS(mp-672) &  19.4 & 82.1 && CdSe(mp-1070) & 6.8 & 42.8  \\
\hline
CdTe(mp-12779) & 4.6 & 42.4 && CuI(mp-569346) &  6.2 & 46.1  \\
\hline
GaP(mp-8882)  &  79.2 & 190.3 && GaSb & 26.8 & 118  \\
& & && (mp-1018059) & & \\
\hline
InAs(mp-1007652) & 18.7 & 93.6 && InP  & 67.6 & 314  \\
& & && (mp-966800) & & \\
\hline
InSb(mp-1007661) & 10.8 & 72 && MgTe(mp-1039) & 11.8 & 75.3  \\
\hline
ZnTe(mp-8884) & 13.1 & 82.6 && Sn(mp-117) & 28.3 & 108.8 \\
\hline
TlMgLa &  9.3 & 54.2 && InN(mp-22205) &  110.1  & 226.3 \\
(mp-962063) & & &&  & &  \\
\hline
AlSb & 131.5 & 417.3 && InN &  108.7 & 235.4\\
(mp-2624) & &  && (mp-20411) \\ 
\hline
AlAs(mp-2172) & 114.7 & 237 && & & \\
\hline
Be$_2$C(mp-1569) & 149.6 & 213 && B$_2$AsP & 433 & 576\\
 &    &  && (mp-1008528) & & \\
\hline
\end{tabular}
\end{center}
\end{table} 

The list of 131 materials used in this work corresponds to some of the most studied and useful insulators and seminconductors for thermal management and thermoelectrics~\cite{MunozRojo2016, MunozRojo2017,thermo1, thermo2, thermo3, thermo4}. From the first-principles calculations it can be observed that most of them fulfil the condition $\ell > \gamma\kappa$, where $\gamma = 1$~nm$\cdot$mK/W (Table~\ref{table_low}). $\gamma$ is just a factor to evaluate the difference in the order of magnitude and units between $\kappa$ and $\ell$. Taking into account this observation, from the data of the studied 131 materials, we use machine learning models to look for materials with some interesting hydrodynamic properties, such as materials with $\ell$ values higher than 1000~nm (similar to C-based materials or BAs) or materials fulfilling the condition $\ell > \gamma\kappa$.

Recently various machine learning techniques have been used to predict the thermal conductivity, like Bayesian optimization~\cite{PhysRevLett.115.205901}, Gaussian process regression~\cite{Abhishek2019} or transfer learning~\cite{Shenghong2020}. In this work we have examined different machine learning models trained using the scikit-learn library in Python. Three types of ensemble learning with regression tree models and neural networks with different combination of layers were considered. The corresponding packages are ExtraTreesRegressor~\cite{extratrees}, RandomForestRegressor~\cite{randomforest}, and GradientBoostingRegressor~\cite{sgb} and XenonPy. Hyperparameters for each type of model were selected using 5-fold cross-validation based on the GridSearchCV package, using in all cases 25$\%$ of data for test. We selected negative mean squared error as the scoring metric. Finally, the model with the lowest root mean squared error (RMSE) on the test data was selected for the high-throughput screening of $\ell$ and $\kappa$.

\begin{figure}[h!]
\includegraphics[width=1.0\linewidth]{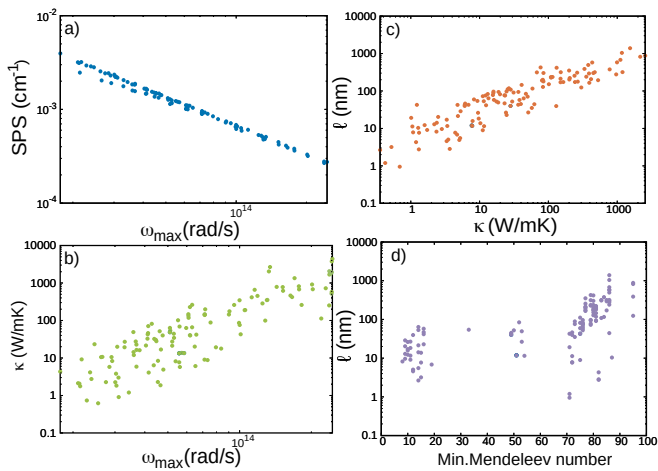}
\caption{Scatter plot of relevant correlations found in the present work for the 131 materials set. \textbf{a)} $SPS$ as function of $\omega_{max}$. \textbf{b)} $\kappa$ as function of $\omega_{max}$. \textbf{c)} $\ell$ as function of $\kappa$. \textbf{d)} $\ell$ as function of Min. Mendeleev number. For $\ell$ and $\kappa$ the median value of the three components have been considered.}
\label{fig:SPS_w_max}
\end{figure}

Considering all the descriptors described in the previous section, those calculated from first-principles and those obtained directly from XenonPy, we have calculated the Pearson's correlation matrix and obtained the corresponding absolute correlation coefficients. By examining the correlation matrix for all the 131 materials we can observe a few interesting things. First of all, we see that the $SPS$ have a correlation of $0.996$ with $w_{max}$ (Fig.~\ref{fig:SPS_w_max}\textbf{a}). From the harmonic point of view, we have also confirmed that $w_{max}$ strongly correlates with $\kappa$ (Fig.~\ref{fig:SPS_w_max}\textbf{b}), with a correlation coefficient of $0.86$, while $v$ and $G_{max}$ with values of $0.78$ and $0.71$ respectively are better than $V$ ($0.53$). Those correlations for the hydrodynamic scale $\ell$ are worse. In the case of $w_{max}$ is $0.59$, while for $v$ and $G_{max}$ are $0.48$ and $0.33$ respectively. Unfortunately the correlation coefficient between $\ell$ and the pseudo thermal conductivity is lower than 0.5 (0.48). On the other hand, the correlation between $\kappa$ and $\ell$ is quite high ($0.88$) (Fig.~\ref{fig:SPS_w_max}\textbf{c}), as both parameters have a strong correlation with the phonon relaxation times according to their definitions~\cite{Torres_book, Torres2018a}. We also observe that $\ell$ strongly correlates with the total relaxation time $\tau$ ($0.92$) and the Normal relaxation time $\tau_N$ ($0.90$), as by definition $\ell = \ell (\tau, \tau_N)$~\cite{Torres_book}. For $\kappa$ these correlations are weaker, $0.72$ and $0.73$ respectively. In addition to the Pearson's correlation coefficients, which accounts for linear correlations, the Maximal Information Coefficient (MIC) has been also calculated, which also consider non-linear correlations. 

\begin{table}[h!] 
\begin{center}
\caption{Top descriptors with MIC correlation with $\ell$ higher than $0.7$.}\label{table_correlation}
\begin{tabular}{lccc}
\hline
\textbf{Descriptor} & \textbf{Correlation with} $\mathbf{\ell}$ \\
\hline
Min. Mendeleev number  &  0.79 \\
\hline
Ave. dipole polarizability  &  0.76 \\
\hline
Max. dipole polarizability  &  0.76 \\
\hline
Max. polarizability  & 0.72 \\
\hline
Min. Allen energy        &    0.71 \\
\hline
Var. Mendeleev number &  0.7 \\
\hline
Min. atomic number    &   0.7 \\
\hline
Min. atomic weight  &  0.7 \\
\hline
Max. Van Der Waals radius (Alvarez)  &  0.7 \\
\hline
\end{tabular}
\end{center}
\end{table} 

The descriptors with a MIC with $\ell$ higher than $0.7$ are summarized in Table~\ref{table_correlation}. Regarding those descriptors we can observe that none of the harmonic or quasi-harmonic descriptors are in the list, and that the highest correlation with  $\ell$ is found for the minimum Mendeleev number ($0.79$) (Fig.~\ref{fig:SPS_w_max}\textbf{d}). While the meaning of this compositional descriptors is self-explained by the names, the Min, Max corresponds to Min/Max values of the descriptor for each unique atom of the compound, and the Ave to the average of the descriptor of all the atoms of the compound. This information provides valuable physical insight that is supported by previous works. As detailed before, the hydrodynamic thermal transport is strongly related to the phonon relaxation times and to several compositional descriptors. Several works have determined analytical expressions for the phonon relaxation times~\cite{Han1993, Han1996, Herring1954, Tamura1983, Klemens1955, Klemens1960, Ward2010}, showing that those magnitudes are directly related with the material density $\rho$, group velocity, and frequency for the case of Normal and Umklapp scattering, in agreement with some of the higher correlated descriptors for $\kappa$. In the case of impurity/mass defect scattering the analytic expresions show the relation to the atomic masses of the atoms of a compound~\cite{Inyushkin2003, Inyushkin2004}. This also agrees with the high correlation of $\kappa$ and $\ell$ with the atomic number and atomic weight. Our calculations have also identified the polarizability as a key factor to determine the intrinsic hydrodynamic effects. A recent work have shown that by strain engineering it is possible to tune the polarizability of a material and modify the thermal conductivity~\cite{PhysRevLett.123.185901}, revealing the high correlation between both magnitudes. In our case the correlation of the Max. polarizability with $\kappa$ is 0.63, while for $\ell$ is 0.76. Therefore it is expected that modifications in the polarizability will have a big influence on $\ell$.



The descriptors for $\kappa$ showing MIC higher than $0.7$ can be found in Table~\ref{table_correlation_kappa}. Note that here we have omitted the $SPS$, showing also a correlation of $0.7$, as for our calculated materials it is almost equivalent to $\omega_{max}$ but it is much more expensive computationally (Fig.~\ref{fig:SPS_w_max}\textbf{a}). In materials with large band gaps this might not be true.

\begin{table}[h!] 
\begin{center}
\caption{Top descriptors with MIC correlation with $\kappa$ higher than $0.7$.}\label{table_correlation_kappa}
\begin{tabular}{lccc}
\hline
\textbf{Descriptor} & \textbf{Correlation with} $\mathbf{\kappa}$ \\
\hline
$w_{max}$       &    0.86 \\
\hline
Min. atomic weight  &  0.79 \\
\hline
Min. atomic number    &   0.79 \\
\hline
$v$       &    0.78 \\
\hline
Ave. heat of formation  &  0.77 \\
\hline
Ave. GS energy  & 0.71 \\
\hline
$G_{max}$       &    0.71 \\
\hline
\end{tabular}
\end{center}
\end{table}


By using the descriptors from Table~\ref{table_correlation} and Table~\ref{table_correlation_kappa}, we first create machine learning models for $\ell$ and $\kappa$ with our known 131 materials calculated from first-principles.

\begin{figure}[h!]
\includegraphics[width=1.0\linewidth]{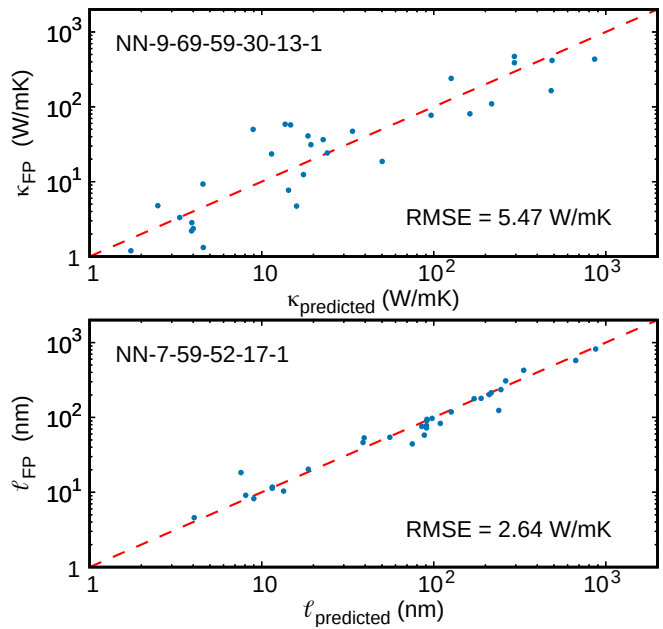}
\caption{Values of $\kappa$ and $\ell$ calculated by first-principles (FP) versus the prediction obtained by machine learning algorithms for the 25$\%$ test set. The straight line represents and ideal exact prediction.}
\label{fig:kappa_ell_predicted}
\end{figure}

A neural network with 6 layers (with 9-69-59-30-13-1 neurons in each layer) has been selected for $\ell$, and a neural network with 5 layers (with 7-59-52-17-1 neurons in each layer) for $\kappa$ as the models with lower RMSE. The best models showed an RMSE value of 2.64~nm for $\ell$ and 5.47~W/mK for $\kappa$ respectively. In Fig.~\ref{fig:kappa_ell_predicted} the first-principles values of $\kappa$ (top) and $\ell$ (bottom) are represented versus the prediction obtained from informatics techniques. It can be observed that the predictions are slightly better for high $\kappa$ values than for the lower ones. Low $\kappa$ materials are highly anharmonic and the present formalism might not be accurate enough. The straight line with slope of 1 represents an exact prediction. The calculated values of $\ell$ and $\kappa$ are averaged over the three axial components.

Now, we proceed to screen all the materials available in the phonon database of the Kyoto University~\cite{Jain2013, ONG2013314, ONG2015209, Togo_MP}. In the phonon database we can find the harmonic information for 10035 materials. From the initial list, we have removed those showing negative phonon bands, reducing the number to 5059. This harmonic information allows us to compute harmonic and quasi-harmonic properties such as $\omega_{max}$, $G_{max}$, $C$, $v$, $SPS$ and $\kappa_{pseudo}^{\alpha, \beta}$, in addition to the material density $\rho$ and volume $V$, for all the 5059 materials.

\begin{table}[h!] 
\begin{center}
\caption{Validation set of materials with $\ell > \gamma \kappa$. $\kappa$ is expressed in W/mK and $\ell$ in nm.}\label{table_4}
\begin{tabular}{lccccccc}
\hline
\textbf{Material} & $\mathbf{\kappa_x}$ & $\mathbf{\kappa_y}$ & $\mathbf{\kappa_z}$ & $\mathbf{\ell_x}$ & $\mathbf{\ell_y}$ & $\mathbf{\ell_z}$\\
\hline
GeAs$_2$(mp-17524) & 8 & 2.6 & 2.4 & 56 & 102.9 & 82.9 \\
\hline
SiP(mp-2798) & 41.9 & 19.4 & 12.9 & 93.5 & 41 & 133.9\\
\hline
BeP$_2$(mp-27148) & 73.5 & 71.5 & 57.5 & 92.9 & 91.4 & 105 \\
\hline
GeO$_2$(mp-470) & 63.6 & 63.6 & 101 & 93.8 & 93.8 & 95.5 \\
\hline
\end{tabular}
\end{center}
\end{table} 


After screening over the  5059 materials, the models have found 147 compounds fulfilling the $\ell > \gamma \kappa$ criteria. From the whole list we have selected 5 materials to check the validity of the predictions from first-principles: GeAs$_2$, SiP, BeP$_2$ and GeO$_2$. The values of thermal conductivity and hydrodynamic scale calculated from first-principles for each axial component can be found in Table~\ref{table_4}. Due to the limited resources, all the 147 materials can not be calculated from first-principles. The full list of materials fulfilling the condition $\ell > \gamma \kappa$ from the phonon database according to our calculations can be found in the APPENDIX.

From Table~\ref{table_4} it can be observed that all the materials fullfil the condition $\ell > \gamma \kappa$ also from first-principles calculations. It is expected that, according to this condition, the materials could be useful for thermal management and/or thermoelectrics. In several works it has been analysed that the dominance of Normal phonon-phonon and electron-phonon scattering in front of Umklapp processes produces a fluid-like behaviour (hydrodynamics) that generates phonon drag and increases the Seebeck coefficient, and consequently an enhancement of the thermoelectric efficiency (ZT)~\cite{Zhou14777,phonondrag1,phonondrag2,phonondrag3,phonondrag4,phonondrag5,phonondrag6}. In the case of SiP, E. Jiang \textit{et. al}~\cite{SiP-2019} showed that in its monolayer form can achieve a ZT$\sim$0.8 and a Seebeck coefficient of 3~mV/K, opening the door to further improvements for nanoscale P-based materials. In the case of BeP$_2$, its bulk ZT is around 1, and in addition several works have highlighted the greatly tunable electronic properties that exhibit in its monolayer and penta-monolayer form~\cite{penta_BeP2-2019, penta_BeP2-2020}. GeAs$_2$ has been reported to be a material with ultralow thermal conductivity and high thermoelectric efficiency~\citep{GeAs2-2017}, showing a Seebeck coefficent higher than 0.3~mV/K with high carrier concentration. Finally, GeO$_2$ has been examined in the last decades as material used to improve the thermoelectric efficiency of other materials by alloying with Cu, SiO$_2$ and CdO to create thin films~\cite{Lucy2006, Snedaker2015, NARUSHIMA2000313} or coating conducting polymers with GeO$_2$ nanoparticles~\cite{pedot}. Unfortunately, the models have not found any new material with high $\ell$ value above 1000~nm.


\begin{figure}[h!]
\includegraphics[width=1.0\linewidth]{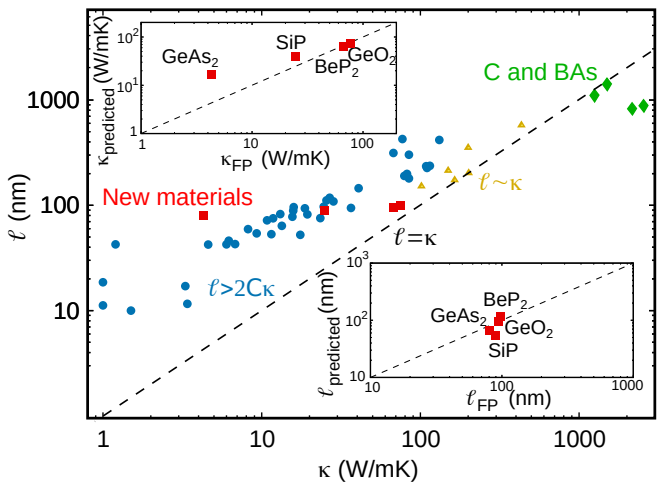}
\caption{Values of $\kappa$ and $\ell$ for the materials corresponding to Table~\ref{table_low} (blue circles), Table~\ref{table_4} (red squares) and C and BN allotropes (green diamonds). The insets show the predicted \textit{VS} the first-principles (FP) calculations for $\kappa$ (top) and $\ell$ (bottom).}
\label{fig:kappa_ell}
\end{figure} 

The results from Table~\ref{table_low} and Table~\ref{table_4} are represented in Fig.~\ref{fig:kappa_ell}, which seems to indicate that thermoelectric materials tend to exhibit a $\ell > \gamma \kappa$ behaviour. The insets show, for the materials of Table~\ref{table_4}, the predicted values by the machine learning models as function of their values calculated from first-principles. The straight line represents the prefect prediction. It can be observed that for low values of $\kappa$ the prediction is worse than for high values. This might be because the present formalism used to calculate the thermal conductivity is not suitable to capture the high anharmonicity of low $\kappa$ materials, affecting the values used to train the models and therefore the predictions. 
In addition, materials with $\ell >$~1000~nm and below the $\ell=\gamma\kappa$ line would be the best candidates to improve heat dissipation in thermal management of electronics.

\section{CONCLUSIONS}

In conclusion, we have used XenonPy's structural and compositional material descriptors together with first-principles calculations to explore the hydrodynamic thermal transport. By using machine learning calculations, we have found that the structural descriptors are not relevant to study the hydrodynamic thermal transport, while there are 9 compositional descriptors with a correlation higher than $0.7$ with $\ell$, the key parameter expressing the scale of phonon hydrodynamics. The calculations have also revealed that diamond and BAs and their allotropes have both high $\ell$. Those materials are the best candidates to identify hydrodynamic signatures from an experimental point of view. The second group of materials that have been considered are those having $\ell > \gamma \kappa$, which seems to be a good indicator to explore new materials to be used for thermal management and thermoelectricity. Here several materials have been found in the original data set, while the machine learning techniques identified 147 materials. From the identified samples, BeP$_2$, SiP, GeAs$_2$ and GeO$_2$ have been calculated from first-principles and used to validate the models, which according to recent works seem to have high thermoelectric performance, in agreement with our original guess. Although these materials have been already reported for this propose, the model found them without knowing of their existence \textit{a priori}. Finally, the full list of values of $\kappa$ and $\ell$ of the materials from the original dataset and the 147 found materials with $\ell > \gamma \kappa$ detailed in the APPENDIX can be useful to look for new compounds for thermal management and thermoelectrics.

\section*{ACKNOWLEDGMENTS}
We acknowledge financial support of The Canon Foundation. This work was partially supported by JSPS KAKENHI (19H00744, 19H01132, 19H50820, 18K18017, 19K14902) and JST CREST (JPMJCR20Q3, JPMJCR19I2, JPMJCR19I1).

\section*{APPENDIX}

In the next table all the values of thermal conductivity $\kappa$ and non-local length $\ell$ for the materials used for the training are displayed. The values that have been already showed in the main text are not included.


\begin{table}[h!]
\caption{Values of $\kappa$ and $\ell$ of the full data set used for training the models. The averaged values of the three components are considered.}\label{table_supp}
\end{table} 
\begin{minipage}{0.5\textwidth}
\begin{center}
\begin{tabular}{ccc}
\hline
\textbf{Material} & $\mathbf{\kappa}$~(W/mK) & $\mathbf{\ell}$~(nm) \\
\hline
Ge(mp-32) &	58.6 & 	116.0 \\
\hline
C(mp-66)	 & 2552.3	 & 880.2 \\
\hline
LiCl(mp-22905) & 	5.1	 & 4.9 \\
\hline
LiBr(mp-23259)	 & 3.5	 & 5.0 \\
\hline
LiI(mp-22899)	 & 3.3	 & 4.6 \\
\hline
NaF(mp-682)	 & 50.9	 & 26.7 \\
\hline
NaCl(mp-22862) & 	9.7	 & 18.3 \\
\hline
NaI(mp-23268)	 & 1.7	 & 9.1 \\
\hline
KF(mp-463)	 & 7.9	 & 15.8 \\
\hline
KCl(mp-23193) & 	9.5	 & 26.1 \\
\hline
KI(mp-22898)	 & 1.6	 & 12.3 \\
\hline
RbF(mp-11718)	 & 2.8	 & 9.6 \\
\hline
RbF(mp-11718)	 & 2.8	 & 9.6 \\
\hline
RbCl(mp-23295) & 	2.2 & 	12.9 \\
\hline
RbBr(mp-22867)	 & 5.3 & 	28.8 \\
\hline
RbI(mp-22903)	 & 2.5	 & 20.3 \\
\hline
AgCl(mp-22922) & 	0.7 & 	0.9 \\
\hline
AgBr(mp-23231) & 	0.4 & 	1.2 \\
\hline
CsF(mp-1784) & 	2.4 & 	8.2 \\
\hline
MgO(mp-1265) & 	57.4 & 	35.2 \\
\hline
CaO(mp-2605)	 & 24.7	 & 22.8 \\
\hline
CaS(mp-1672) & 	38.0	 & 57.6 \\
\hline
CaSe(mp-1415)	 & 23.1 & 	46.7 \\
\hline
CaTe(mp-1519)	 & 12.4	 & 41.6 \\
\hline
SrO(mp-2472)	 & 10.3	 & 13.8 \\
\hline
BaO(mp-1342) & 	5.2 & 	5.5 \\
\hline
\end{tabular}
\end{center}
\end{minipage} \hfill

\begin{minipage}{0.5\textwidth}
\begin{center}
\begin{tabular}{ccc}
\hline
BaS(mp-1500)	 & 4.8 & 	21.4 \\
\hline
CdO(mp-1132)	 & 11.1 & 	8.9 \\
\hline
PbS(mp-21276) & 	3.7 & 	2.8 \\
\hline
PbSe(mp-2201)	 & 1.3 & 	2.8 \\
\hline
PbTe(mp-19717)	 & 1.2 & 	4.3 \\
\hline
AlN(mp-1700)	 & 267.2	 & 177.9 \\
\hline
BAs(mp-10044)	 & 1537.8	 & 1389.0 \\
\hline
BeO(mp-1778) &  527.5 & 	194.2 \\
\hline
BeSe(mp-1541) & 	239.6	 & 255.2 \\
\hline
BeTe(mp-252)	 & 200.0	 & 351.7 \\
\hline
BN(mp-1639)	 & 1192.7 & 	319.1 \\
\hline
BP(mp-1479)	 & 449.8	 & 301.3 \\
\hline
CuCl(mp-22914)	 & 1.3	 & 7.7 \\
\hline
CuI(mp-22895)	 & 6.1 & 	39.8 \\
\hline
GaN(mp-830) & 	201.4 & 	202.4 \\
\hline
SiC(mp-8062)	 & 471.5 & 	317.4 \\
\hline
ZnO(mp-1986)	 & 49.9 & 	70.6 \\
\hline
AlAs(mp-8881) & 	83.0	 & 199.0 \\
\hline
CdO(mp-1132)	 & 11.1 & 	8.9 \\
\hline
PbS(mp-21276) & 	3.7 & 	2.8 \\
\hline
PbSe(mp-2201)	 & 1.3 & 	2.8 \\
\hline
PbTe(mp-19717)	 & 1.2 & 	4.3 \\
\hline
AlN(mp-1700)	 & 267.2	 & 177.9 \\
\hline
BAs(mp-10044)	 & 1537.8	 & 1389.0 \\
\hline
BeO(mp-1778) &  527.5 & 	194.2 \\
\hline
BeSe(mp-1541) & 	239.6	 & 255.2 \\
\hline
BeTe(mp-252)	 & 200.0	 & 351.7 \\
\hline
BN(mp-1639)	 & 1192.7 & 	319.1 \\
\hline
BP(mp-1479)	 & 449.8	 & 301.3 \\
\hline
CuCl(mp-22914)	 & 1.3	 & 7.7 \\
\hline
CuI(mp-22895)	 & 6.1 & 	39.8 \\
\hline
GaN(mp-830) & 	201.4 & 	202.4 \\
\hline
SiC(mp-8062)	 & 471.5 & 	317.4 \\
\hline
ZnO(mp-1986)	 & 49.9 & 	70.6 \\
\hline
AlAs(mp-8881) & 	83.0	 & 199.0 \\
\hline
AlN(mp-661)	 & 243.4	 & 173.2 \\
\hline
BAs(mp-984718)	 & 1169.7	 & 1043.6 \\
\hline
BeO(mp-2542)	 & 392.9	 & 165.5 \\
\hline
BN(mp-2653) & 	965.8 & 	277.1 \\
\hline
BP(mp-1008559)	 & 355.5	 & 240.3 \\
\hline
CdS(mp-672)	 & 19.8 & 	81.6 \\
\hline
CdTe(mp-12779) & 	4.8	 & 42.5 \\
\hline
CuCl(mp-1184046) & 	1.1 & 	7.8 \\
\hline
GaAs(mp-8883)	 & 31.2	 & 91.0 \\
\hline
GaN(mp-804)	 & 204.4	 & 222.0 \\
\hline
GaP(mp-8882) & 	80.7	 & 190.8 \\
\hline
SiC(mp-7140)	 & 389.0	 & 293.3 \\
\hline
ZnO(mp-2133)	 & 43.7 & 63.5 \\
\hline
ZnS(mp-560588)	 & 23.0	 & 76.9 \\
\hline
ZnSe(mp-380)	 & 16.1 & 	57.9 \\
\hline
AgSbBa(mp-984720) & 	0.4	 & 2.7 \\
\hline
CoAsZr(mp-961689) & 	36.5 & 	42.8 \\
\hline
CoSbZr(mp-22377) & 	39.8	 & 44.9 \\
\hline
NiScSb(mp-3432) & 	4.7 & 6.8 \\
\hline
BeO(mp-1778) &  527.5 & 	194.2 \\
\hline
BeSe(mp-1541) & 	239.6	 & 255.2 \\
\hline
BeTe(mp-252)	 & 200.0	 & 351.7 \\
\hline
BN(mp-1639)	 & 1192.7 & 	319.1 \\
\hline
BP(mp-1479)	 & 449.8	 & 301.3 \\
\hline
\end{tabular}
\end{center}
\end{minipage} \hfill

\begin{minipage}{0.5\textwidth}
\begin{center}
\begin{tabular}{ccc}
\hline
CuCl(mp-22914)	 & 1.3	 & 7.7 \\
\hline
CuI(mp-22895)	 & 6.1 & 	39.8 \\
\hline
GaN(mp-830) & 	201.4 & 	202.4 \\
\hline
SiC(mp-8062)	 & 471.5 & 	317.4 \\
\hline
ZnO(mp-1986)	 & 49.9 & 	70.6 \\
\hline
AlAs(mp-8881) & 	83.0	 & 199.0 \\
\hline
CdO(mp-1132)	 & 11.1 & 	8.9 \\
\hline
PbS(mp-21276) & 	3.7 & 	2.8 \\
\hline
NiSnTi(mp-924130)	 & 7.7	 & 12.0 \\
\hline
NiSnTi(mp-623646) & 	7.7	 & 12.0 \\
\hline
NiSnTi(mp-22782) & 	7.7 & 	12.0 \\
\hline
NiSnZr(mp-924129) & 	28.1	 & 39.9 \\
\hline
NiSnZr(mp-30806)	 & 28.1	 & 39.9 \\
\hline
PdSrTe(mp-961663)	 & 0.5 & 	3.2 \\
\hline
PtGaTa(mp-961677) & 	47.1	 & 84.2 \\
\hline
PtGeTi(mp-1008680) & 	7.8	 & 11.7 \\
\hline
PtGeTi(mp-961671) & 	7.8	 & 11.7 \\
\hline
PtNbIn(mp-961708) & 	18.6	 & 49.4 \\
\hline
RhNbSi(mp-1100400) & 	24.2 & 	26.5 \\
\hline
RhSbHf(mp-10367)	 & 29.8	 & 52.3 \\
\hline
RuAsV(mp-1100404)	 & 11.6	 & 11.5 \\
\hline
SiSrCd(mp-962076)	 & 4.4	 & 7.7 \\
\hline
SnSrBa(mp-962062) & 	2.4	 & 14.2 \\
\hline
GeC(mp-1002164)	 & 776.3	 & 376.9 \\
\hline
Lonsdaleite(mp-47)	 & 2159.6 & 	821.4 \\
\hline
BC$_2$N(mp-30148) & 	1112.1	 & 679.2 \\
\hline
SiSn(mp-1009813) & 	151.1	 & 317.0 \\
\hline
BSb(mp-997618) & 	379.5	 & 494.7 \\
\hline
B$_2$AsP(mp-1008528)	 & 433.5 & 	576.6 \\
\hline
C$_3$N$_4$(mp-571653) & 	334.4	 & 387.9 \\
\hline
BC$_2$N(mp-629458) & 993.8	 & 581.5 \\
\hline
BeCN$_2$(mp-15703) & 	416.3 & 	207.2 \\
\hline
C$_3$N$_4$(mp-2852) & 300.6	 & 124.1 \\
\hline
SnC(mp-1009820)	 & 261.7 & 	308.1 \\
\hline
B(mp-160)	 & 62.5	 & 44.3 \\
\hline
LiBC(mp-9244) & 	130.0	 & 39.6 \\
\hline
\end{tabular}
\end{center}
\end{minipage} \hfill
\newline
\newline
The full list of materials identified by the machine learning models fulfilling the condition $\ell > \gamma \kappa$ are:
\newline
\newline
Se(mp-147), Sm$_2$O$_3$(mp-218), YbSe(mp-(286), B$_2$O$_3$(mp-306), MgP$_4$(mp-384), Sc$_2$S$_3$(mp-401), CdP$_2$(mp-402), AlF$_3$(mp-468),GeO$_2$(mp-470), NiP$_2$(mp-486), Ga$_2$S3(mp-539), TeO$_2$(mp-557), GaF$_3$(mp-588), B$_6$As(mp-624), SnSe$_2$(mp-665), GeSe(mp-700), P$_2$Pt(mp-730), Li$_3$P(mp-736), Li$_3$As(mp-757), Ga$_2$O3(mp-886), As$_2$Se$_3$(mp-909), CuP$_2$(mp-927), GeTe(mp-938), NbZn$_3$(mp-953), K$_2$O(mp-971), SrF$_2$(mp-981), BaF$_2$(mp-1029), SrS(mp-1087), Al$_2$O$_3$(mp-1143), ThS$_2$(mp-1146), YbO(mp-1216), MgS(mp-1315), Ga$_2$Se$_3$(mp-1340), ZnP$_2$(mp-1392), Rb$_2$O(mp-1394), Sn$_2$S$_3$(mp-1509), As$_2$O$_3$(mp-1581), Al$_4$C$_3$(mp-1591), Tm$_2$O3(mp-1767), YbTe(mp-1779), SiF$_4$(mp-1818), YbS(mp-1820), SiAs(mp-1863), ZnF$_2$(mp-1873), SrTe(mp-1958), Mg$_3$As$_2$(mp-1990), Li$_3$Sb(mp-2074), Rb$_2$P$_3$(mp-2079), Si5Ir$_3$(mp-2084), KGe(mp-2146), Er$_2$S3(mp-2234), GeS(mp-2242), Tm$_2$S$_3$(mp-2309), Dy$_2$O$_3$(mp-2345), Ga$_2$Te$_5$(mp-2371), NaSi(mp-2402), P$_4$Se$_5$(mp-2447), Al$_2$S$_3$(mp-2654), CaF$_2$(mp-2741), SrSe(mp-2758), SiP(mp-2798), Lu$_2$S$_3$(mp-2826), InF$_3$(mp-6949), Be$_3$N$_2$(mp-6977), Rb$_2$Se$_3$(mp-7447), GeF$_2$(mp-7595), K$_2$Se$_3$(mp-7670), ThSe$_3$(mp-7951), PF$_5$(mp-8511), AsF$_5$(mp-8723), Al$_2$Te$_5$(mp-9254), GeAs(mp-9548), ZnS(mp-9946), MgSe(mp-10760), RbSn(mp-11054), Sr$_3$As$_4$(mp-15339), Ge$_3$Os$_2$(mp-16610), GeAs$_2$(mp-17524), InS(mp-19795), InSe(mp-20485), Si$_3$Ru$_2$(mp-22192), In$_2$O$_3$(mp-22323), CsCl(mp-22865), YbBr$_2$(mp-22882), XeF$_2$(mp-22885), CaBr$_2$(mp-22888), ZnCl$_2$(mp-22889), CaCl$_2$(mp-22904), CsBr(mp-22906), PtBr$_3$(mp-23165), IO$_2$(mp-23170), SrI$_2$(mp-23181), BaCl$_2$(mp-23199), InI(mp-23202), MgI$_2$(mp-23205), SrCl$_2$(mp-23209), MgCl$_2$(mp-23210), AsI$_3$(mp-23218), YbCl$_2$(mp-23220), BaI$_2$(mp-23260), BeCl$_2$(mp-23267), BeP$_2$(mp-27148), ZnI$_2$(mp-27161), BaBr$_2$(mp-27456), Te$_3$Cl$_2$(mp-27628), ThI$_4$(mp-27697), PdI$_2$(mp-27747), GeI$_2$(mp-27922), CdI$_2$(mp-28248), P$_2$Pd(mp-28266), PtI$_2$(mp-28319), Pt$_3$I$_8$(mp-28320), GaBr$_2$(mp-28384), Br$_2$O(mp-28460), In$_7$Cl$_9$(mp-28730), Ca$_5$P$_8$(mp-28879), Cl$_2$O(mp-29537), NaGe(mp-29657), Tl$_2$Te$_3$(mp-29711), LaAs$_2$(mp-29815), CaI$_2$(mp-30031), MgBr$_2$(mp-30034), BeI$_2$(mp-30140), GaCl$_3$(mp-30952), Si$_9$Te$_8$(mp-31135), NbI$_5$(mp-31487), Tm$_2$Se$_3$(mp-32850), AuBr(mp-505366), Cs$_2$S(mp-540703), KSn(mp-542374), ThCl$_4$(mp-567431), SrBr$_2$(mp-567744), HfI$_4$(mp-569059), SiB$_3$(mp-569128), TeI(mp-569766), ThBr$_4$(mp-570229), YbI$_2$(mp-570418), SeBr(mp-570589), Ga$_3$Os(mp-570844), ZrI$_4$(mp-571235), Ga$_2$I$_3$(mp-636675), In$_7$Te$_10$(mp-669311), Lu$_2$Se$_3$(mp-673650), Sc$_2$Se$_3$(mp-684690), CsSi(mp-866482), SiGe(mp-978534).

\bibliography{references}

\end{document}